# Embodied Virtual Reality Feedback Reshapes Neural Representations to Support Continuous Three-Dimensional Motor Imagery Decoding

**Niall McShane[1]*, Attila Korik[2], Karl McCreadie[3], Naomi Du Bois[2], Darryl Charles[1], Damien Coyle[2,3]**

[1] Performance and Practice-led Research in Arts, Culture, and Creative Technologies Research Centre, School of Arts and Humanities, Ulster University, Derry/Londonderry, United Kingdom

[2] The Bath Institute for the Augmented Human, University of Bath, Bath, United Kingdom

[3] Intelligent Systems Research Centre, School of Computing, Engineering and Intelligent Systems, Ulster University, Derry/Londonderry, United Kingdom

***Correspondence:**
Niall McShane
n.mcshane@ulster.ac.uk

**Abstract**

Continuous brain-computer interfaces (BCIs) that decode motion trajectories from imagined movement offer intuitive motor control, yet how feedback modality and longitudinal training shape neural representations and decoding performance remains poorly understood. Here we present the first systematic investigation of embodied virtual reality (VR) feedback during real-time three-dimensional virtual limb control driven by motor imagery, across ten longitudinal sessions in ten participants. Decoding performance was evaluated using three complementary strategies spanning actual online performance (Fixed Decoder Generalisation, FDG), periodic retraining (Sequential Adaptive Training, SAT), and within-session upper-bound estimation (Within-Session Reconstruction, WSR), enabling dissociation of user adaptation effects from decoder optimisation. A CNN-LSTM decoder optimised via asynchronous successive halving achieved within-session imagined movement correlations of $r = 0.762$ under VR and $r = 0.672$ under screen feedback (WSR). VR significantly outperformed screen feedback across all three evaluation strategies and movement dimensions, with mean correlation improvements of 8.9–13.0% (all $p \leq 0.002$, $d = 1.42$–$2.05$). Critically, this advantage persisted under fixed decoders without retraining, demonstrating that embodied VR feedback elicits inherently more decodable and generalisable neural representations rather than simply improving decoder fit. Linear mixed-effects modelling confirmed robust main effects of feedback modality and movement axis with no interaction, indicating a consistent VR advantage across all movement directions. Neurophysiologically, VR produced stronger sensorimotor-parietal desynchronisation and enhanced motor-frontal functional connectivity, with pervasive engagement of the anterior insula across all frequency bands and increased superior parietal lobule coupling, paralleling neural patterns associated with real movement execution. These findings establish embodied spatial feedback as a key design principle for next-generation continuous BCIs targeting intuitive motor control and neurorehabilitation.

## 1 Introduction

### 1.1 Background and Motivation

Electroencephalogram (EEG)-based brain-computer interfaces (BCIs) have been investigated for communication and control applications for people with a physical disability [1] and in neurorehabilitation [2], assessment in disorders of consciousness [3,4], assistive software applications [5], entertainment, and computer games [6], [7]. Motor imagery (MI) BCIs, in which users imagine rather than execute movements to modulate their sensorimotor neural activity, can be used to generate continuous control outputs across multiple degrees of freedom, including 3D trajectory control. A subset of such systems, referred to here as motion trajectory decoding BCIs (MTD-BCIs), decode neural correlates of limb movement, whether executed or imagined, to produce continuous 3D control outputs; in online control phases, decoding targets the neural correlates of imagined rather than executed movement, providing more intuitive control than conventional discrete classification approaches [8–13]. Whilst traditional linear regression (LR) and multiple linear regression (mLR) models have been widely used, they assume linear relationships between neural activity and movement parameters, often failing to capture the nonlinear dynamics inherent to EEG signals [9,10,12,14]. Given these complexities, advanced deep learning (DL) methods now provide more powerful nonlinear mapping, yielding significant gains in decoding accuracy (DA) [11,15–18].

Convolutional neural network-long short-term memory (CNN-LSTM) architectures, which combine convolutional spatial feature extraction with recurrent temporal modelling, have proven particularly effective for continuous trajectory decoding. Pancholi et al. (2022) showed that CNN-LSTM models outperform mLR for continuous hand-trajectory decoding, achieving maximum $r = 0.67$ using source-localised EEG features [15], while Jain and Kumar (2023) demonstrated subject-independent 3D hand kinematics reconstruction using a CNN-LSTM framework [19]. LSTM-based approaches have also improved reaching trajectory prediction from magnetoencephalography signals [20], and end-to-end LSTM architectures have been applied to MI classification tasks [21,22]. More recently, Li et al. (2024) combined feature selection with nonlinear regression methods for continuous upper limb trajectory decoding from EEG, achieving $r = 0.511$ and surpassing traditional ridge regression and mLR approaches [23]. Interpretable CNNs such as MS-Sinc-ShallowNet have further improved 2D decoding while revealing delta-band contributions as particularly informative for hand kinematics [16,24]. For context, reported correlation values across established MTD approaches typically range from $r = 0.15$ to 0.70 depending on movement type, task complexity, and decoding strategy [8,12,23,25], with higher accuracies generally observed for executed compared to imagined movements and for faster movement speeds [26]. Despite these advances, most MTD studies remain restricted to offline analyses or 2D screen feedback, leaving real-time learning and user adaptation underexplored. Yet effective continuous BCI control depends not only on algorithmic accuracy but also on how feedback design shapes neural dynamics.

Recent theoretical work has formalised this challenge. Wolpaw (2025) argues that a BCI does not simply decode brain activity but enables its user to acquire a new skill produced by a "synthetic heksor" (a network of neurons, synapses, and BCI software that must adapt continuously to maintain the key features of that skill) [27,28]. Critically, improved signal analysis increases mean DA but does not reduce the session-to-session variability that limits practical reliability, because the synthetic heksor must negotiate with the many natural heksors (skill-producing networks) with which it shares neural resources. This framework highlights two underexplored requirements for reliable continuous BCIs: effective co-adaptation between the user's central nervous system and the decoder software, and sensory feedback rich enough to guide ongoing neural adaptation, both of which are addressed in the present study.

## 1.2 The Role of Feedback and Neural Connectivity

Feedback modality strongly influences brain connectivity during motor imagery (MI) tasks [29–31]. Functional connectivity (FC) provides a broader view of coordinated neural activity than analyses limited to local sensorimotor rhythms (SMRs) [32]. This network-level perspective is particularly relevant given that FC patterns differ substantially between resting and task states, with motor tasks reorganising sensorimotor and visual networks relative to baseline activity [32]. Leeuwis et al. (2021) demonstrated that MI-BCI performance is closely related to connectivity strength, with high-performing users showing increased alpha-band coupling between motor and visuospatial processing regions [30]. They further reported that this enhanced inter-regional connectivity, particularly within right-hemisphere networks which exhibit alpha modulations, corresponded to improved control performance [30]. Complementary topographical analyses have further shown that mu, beta, and low-gamma oscillations encode 3D hand kinematics with characteristic spatial distributions over central and parietal regions, suggesting that both inter-regional connectivity and local activation patterns are relevant to understanding the cortical dynamics underlying MI-based trajectory control [8].

The modality through which feedback is delivered appears to shape these connectivity patterns differentially. Comparative work on Virtual Reality (VR) versus screen (i.e. computer monitor) feedback revealed distinct network differences: EEG analyses using partial directed coherence identified greater alpha-band efficiency in prefrontal regions under first-person VR, along with enhanced motor-to-frontal connectivity across alpha and beta bands [33]. These findings suggest immersive VR feedback promotes stronger frontal engagement and integration of motor signals, which improve DA.

VR environments provide an effective means of delivering spatially aligned feedback, allowing users to perceive and control a body-referenced effector within an egocentric 3D space. Spatial feedback using virtual limbs or hands is consistent with established mechanisms of visuo-proprioceptive embodiment demonstrated in both real-world contexts and earlier VR research [34,35]. This embodied spatial feedback engages broader sensorimotor networks and strengthens frontoparietal circuits compared with conventional screen feedback [36,37]. EEG-fNIRS studies further show that VR visual cues increase overall brain network integration during MI, with higher mean degree and clustering coefficient after VR-based practice [36]. Directed FC analyses comparing movement imagery, execution, and observation have revealed distinct patterns of information flow among motor regions, suggesting that the cognitive state induced by different task contexts fundamentally alters network organisation [38]. fMRI studies of VR-based motor imagery and observation tasks similarly demonstrate enhanced activation in motor-related networks, supporting the hypothesis that immersive environments engage neural systems more similarly to actual movement execution [37]. Multimodal approaches that combine visual and electro-tactile feedback delivered within VR environments have enhanced communication between motor and higher-order cognitive regions while reducing sensory cortical workload [29]. These network reorganisations suggest that VR feedback promotes more efficient interaction between motor command and cognitive control systems.

Although VR feedback has been shown to increase embodiment and presence in classical, discrete MI-BCIs involving the discrimination of two separate limbs [39–41], its specific impact on continuous 3D trajectory decoding during 3D motor imagery remains unclear. To date, most MTD research has relied on abstract or non-embodied feedback paradigms rather than leveraging spatially congruent, embodied feedback to enhance DA and influence neural dynamics during real-time control. Previous MTD-BCI studies have primarily focused on short-term, proof-of-concept experiments, typically limited to a single session or a few repetitions, aimed at demonstrating algorithmic feasibility rather than user adaptation or neural plasticity [8,16,26,42,43]. Consequently, little is known about how users learn and stabilise continuous control through extended interaction with real-time feedback in MI-based MTD-BCIs.

Training protocols strongly influence BCI performance but remain underexplored in continuous 3D MTD contexts. Multi-session studies in MI-BCI show that the structure and sequencing of feedback modalities shape user adaptation and engagement, with gradual increases in feedback complexity enhancing motivation and learning [44]. Across extended training, adaptation often follows non-linear trajectories, with plateaus and inter-session variability particularly evident when fixed decoders cannot accommodate evolving neural strategies [45–47]. Despite their importance, most training protocols remain designed for discrete MI tasks, with little attention to continuous, embodied motion decoding in 3D space during 3D motor imagery. Previous work has also highlighted design limitations in 3D MTD paradigms, where suboptimal target configurations can reduce separability and compromise decoder training due to conflated velocity signatures [10,11]. Eye gaze integration represents another design consideration that has received limited attention. Although BCI protocols typically constrain gaze to prevent signal contamination, natural reaching movements follow a 'look-then-reach' pattern requiring visual target acquisition before movement initiation [48]. Accommodating this natural visuomotor coordination may enhance the ecological validity of continuous MTD systems. Building on these observations, the present study provides a systematic examination of how VR feedback, longitudinal training with graduated assistance reduction, and highly optimised deep learning models jointly influence MTD-BCI decoding accuracy, user adaptation, and cortical network dynamics during real-time 3D virtual limb trajectory control.

## 2 Methods

### 2.1 Experimental Protocol

Ten right-handed, able-bodied participants (9 male, 1 female; mean age 21.2 ± 1.3 years, range 20–24), naïve to BCI use, completed ten sessions each lasting 90–120 minutes. Each session contained 256 trials (128 executed, 128 imagined), producing 2,560 trials per participant across ten sessions (25,600 total). Session 1 included orientation and practice to standardise task timing; subsequent sessions followed a consistent protocol of EEG preparation, calibration checks, and runs. A within-participant, repeated-measures design was used. Visual feedback alternated between screen and VR on a session-by-session basis, with starting modality counterbalanced across participants (odd-numbered participants beginning with VR, even-numbered with screen), such that each participant experienced a fully alternating sequence across all ten sessions. With N = 10 and $\alpha = 0.05$, an a priori power analysis for a two-tailed paired t-test (G*Power 3.1; [49]) indicated that this sample size was sufficient to detect large within-subject effects (Cohen's d = 1.0, 1 - $\beta$ = 0.80), consistent with effect sizes reported in comparable MI-BCI studies [39,41].

Prior to online control, modality-specific decoders were trained during offline sessions (S1–S2) for each feedback condition separately. EEG recorded during motor imagery trials was used as the decoder input, with velocity labels derived from executed movement kinematics as the decoding target, and 100% trajectory assistance applied to imagined trials to ensure consistent visual feedback throughout training. The best-performing model per modality was selected for the subsequent online sessions (S3–S10); for each feedback modality, the decoder trained on the first session's EEG data was then used for 3D motor imagery decoding across the four online sessions. Trajectory assistance was applied to the virtual hand feedback during online sessions to support learning and provide slightly positively biased feedback. Assistance was initially set at 65% in online sessions S3–S4 and reduced every two sessions to 60%, 50%, and 40% in the final sessions S9–S10.

### 2.2 Experimental Setup and Trial Timing

Participants controlled an embodied virtual hand with matched kinematics in both visual feedback modalities (Figure 1D). In the VR condition, feedback was first-person and stereoscopic; in the screen condition, the same hand and motions were rendered as a 2D projection on a monitor. Movement onset and return-to-rest were cued by a 440 Hz tone (64 ms) and visual target outline. In the VR condition, audio cues were spatialised to the target location using Unity's Audio Spatializer SDK, which applies head-related transfer functions (HRTFs) to simulate the direction and distance of sound sources relative to the listener, providing egocentric auditory localisation consistent with the 3D spatial layout of the task. The use of spatialised rather than non-spatialised audio is relevant in the BCI context, as McCreadie et al. demonstrated that the modality and spatial configuration of auditory feedback can influence sensorimotor BCI performance [50], and may have further reinforced the embodied spatial frame of reference during VR trials. In the screen condition, the same tones were presented via stereo monitor speakers without spatialisation.

Targets were presented at four spatial positions arranged around a central fixation point directly in front of the right shoulder. To prevent trajectory conflation observed in standard cross arrays [11], i.e., arranged perpendicularly, targets were offset in the x, y, and z planes to increase separability of velocity signatures and promote more natural reach biomechanics (Figure 1A). This configuration improved the distinctiveness of 3D velocity patterns and the quality of training data for decoding.

Participants completed alternating runs of executed and imagined reaches, organised into eight runs per session (four executed, four imagined) and subdivided into 16-trial blocks with randomised target order (Figure 1B-C). Eye movements were minimised and blinks restricted to rest phases to reduce artefacts. Consistent with natural reach behaviour [48], participants fixated on the central cross during rest and shifted gaze to the cued target during the pre-movement indication period. Participants self-initiated each block of trials, triggering a 30 s countdown before the first trial. Each trial lasted 7.6 s and followed four distinct phases (Figure 1D): a Rest phase (0.0–2.5 s) for baseline fixation; an Indication phase (2.5–4.1 s) with a 1.6 s pre-cue line extending from the fixation cross to the upcoming target; a Target phase (4.1–6.6 s) where a 440 Hz tone (64 ms) signalled movement onset, the target outline illuminated and expanded 1.5× over 250 ms and an animated trajectory line appeared from the virtual wrist to the target to guide movement; and a Reset phase (6.6–7.6 s) cued by a second tone, signalling returning the virtual hand to its resting position. This structured sequence ensured synchronised gaze behaviour, clear visual-auditory cues, and precise alignment of EEG and behavioural data during both executed and imagined control.

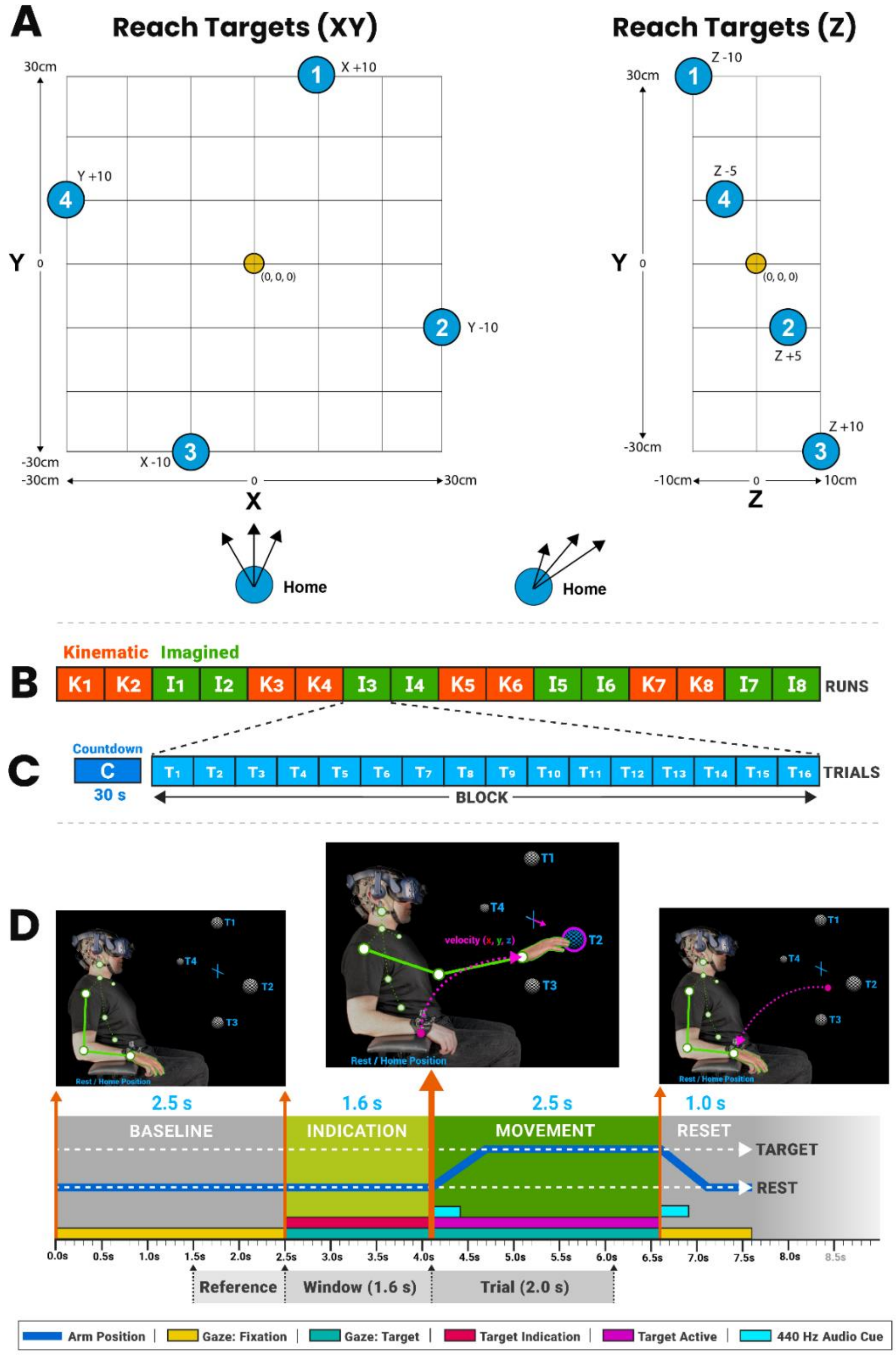

*Figure 1: Target configuration and experimental setup. (A) Spatially offset 3D target array to optimise trajectory separability and reach biomechanics. (B-C) Session structure showing alternating executed and imagined runs, divided into 16-trial blocks with randomised targets. (D) Trial timing sequence comprising rest, indication, target presentation, and reset phases.*

### 2.3 Hardware and Data Collection

EEG was recorded using a g.tec g.Nautilus wireless system equipped with 32 g.SCARABEO gel-based electrodes plus ground and reference connections [51]. A 32-channel montage (C3, C4, AF3, T7, F7, F3, FZ, F4, T8, FC5, FC1, FC2, FC6, AF4, PO7, PO8, CZ, F8, CP5, CP1, CP2, CP6, P7, P3, PZ, P4, P8, CP3, PO3, PO4, OZ, CP4) was used to achieve complete scalp coverage (Figure 2A). To improve efficiency and DL training, online decoding and analysis was limited to a 17-channel montage (F3, FZ, F4, FC5, FC1, FC2, FC6, C3, CZ, C4, CP5, CP1, CP2, CP6, P3, PZ, P4), centred over sensorimotor regions (Figure 2B). This same montage was used for evaluating DA to ensure consistency between online performance and offline analyses.

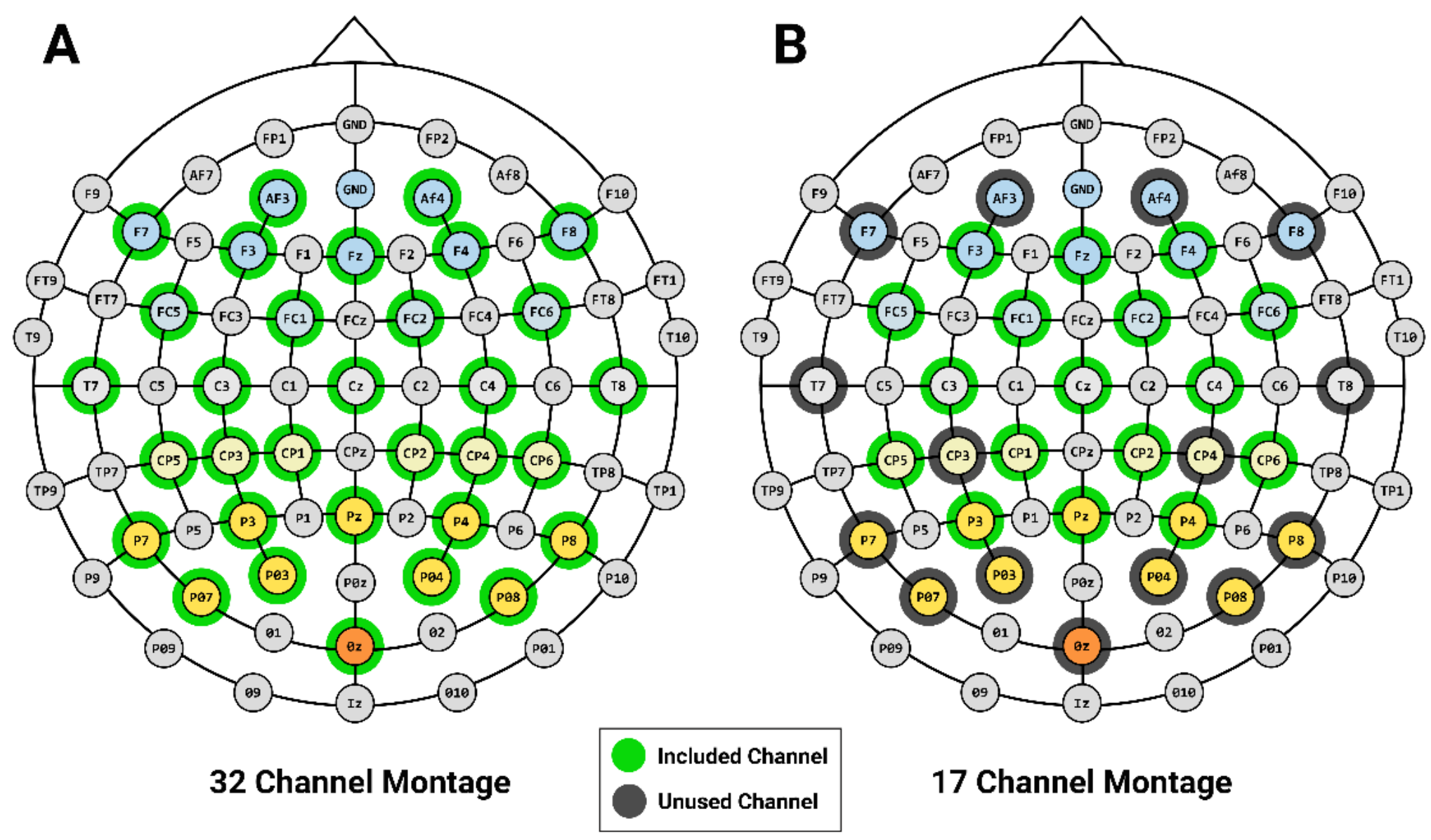

*Figure 2: EEG Montages – The 17 sensorimotor-centred channels (Online Montage) were used for real-time MTD-BCI decoding and online/offline DA analysis. 32-channel coverage was used for FC and topographical analysis to assess broader cortical network dynamics (Functional Connectivity analysis Montage).*

Kinematic motion data was captured within the paradigm software at 60 Hz from a wrist-mounted Vive tracker using four SteamVR base station 2.0 units angled downwards ~45° from opposing room corners at a height of 264 cms. Visual feedback was presented either on a 42.5-inch Dell P4317Q monitor (3840 × 2160 px, 60 Hz) or via a Vive Pro Eye HMD (1440 × 1600 px per eye, 90 Hz). Gaze monitoring was performed using modality-matched eye-tracking systems: a Tobii Eye Tracker 5 (133 Hz sampling rate, 40° × 40° field of view) externally mounted on the screen, and the integrated Tobii eye-tracking system within the Vive Pro Eye HMD (120 Hz, 110° field of view), ensuring comparable temporal resolution and spatial coverage across both feedback conditions to support central fixation monitoring during task periods.

Three-dimensional position samples (x, y, z) were numerically differentiated to obtain velocity using a first-order finite difference approximation at each time step [Eq. (1)], where $p_i = (x_i, y_i, z_i)$ is the position vector at sample $i$, and $v_i$is the corresponding instantaneous velocity vector. This discrete formulation approximates $dp/dt$by dividing the positional displacement between consecutive samples by the sampling interval ($\Delta t \approx$ 16.67 ms at 60 Hz). Velocity estimates were temporally aligned with EEG triggers for subsequent decoding analysis.

$$v_i = \frac{p_i - p_{i-1}}{t_i - t_{i-1}} \quad (1)$$

### 2.4 CNN-LSTM Decoder

An event-related spectral perturbation (ERSP) fed CNN-LSTM deep learning architecture was implemented to decode continuous 3D limb velocity from EEG. ERSP is a time-frequency analysis method that quantifies event-related changes in spectral power relative to a baseline period, providing a rich representation of motor-related neural dynamics suitable for CNN-based feature extraction [52]. CNNs enable hierarchical spatial feature learning from structured inputs, while LSTM networks are designed to capture temporal dependencies in sequential data. The combined ERSP-CNN-LSTM framework has previously been shown to successfully decode continuous lower-limb stepping velocity from EEG using ERSP representations [18]. The model therefore integrates convolutional feature extraction with recurrent temporal modelling to capture spatial-temporal dependencies in ERSP representations of the EEG signal (Figure 3A-B). ERSP images were generated for each of the 17 EEG channels, with frequency (0–40 Hz, 1 Hz resolution) on the vertical axis and time on the horizontal axis. Time windows began 640 ms before trial onset and extended across the movement period using 16 ms observation steps. Each 2D ERSP image (40 × 40 pixels) was normalised to a resting reference period and stacked to form a channel-wise input volume for the CNN (Figure 3A). The CNN comprised three 2D convolutional layers (3 × 3 kernels, 32 filters in the first layer) with max-pooling and dropout regularisation before 1D flattening (Figure 3B). Output features were passed to two stacked LSTM layers (50 units in the first, dropout = 0.25) to model temporal dependencies and maintain context across sequential windows. The *tanh* activation function was used for hidden layers, with a dense output layer producing the decoded x-, y-, and z-velocity components. Separate single-output models were also trained per velocity dimension for comparative analysis.

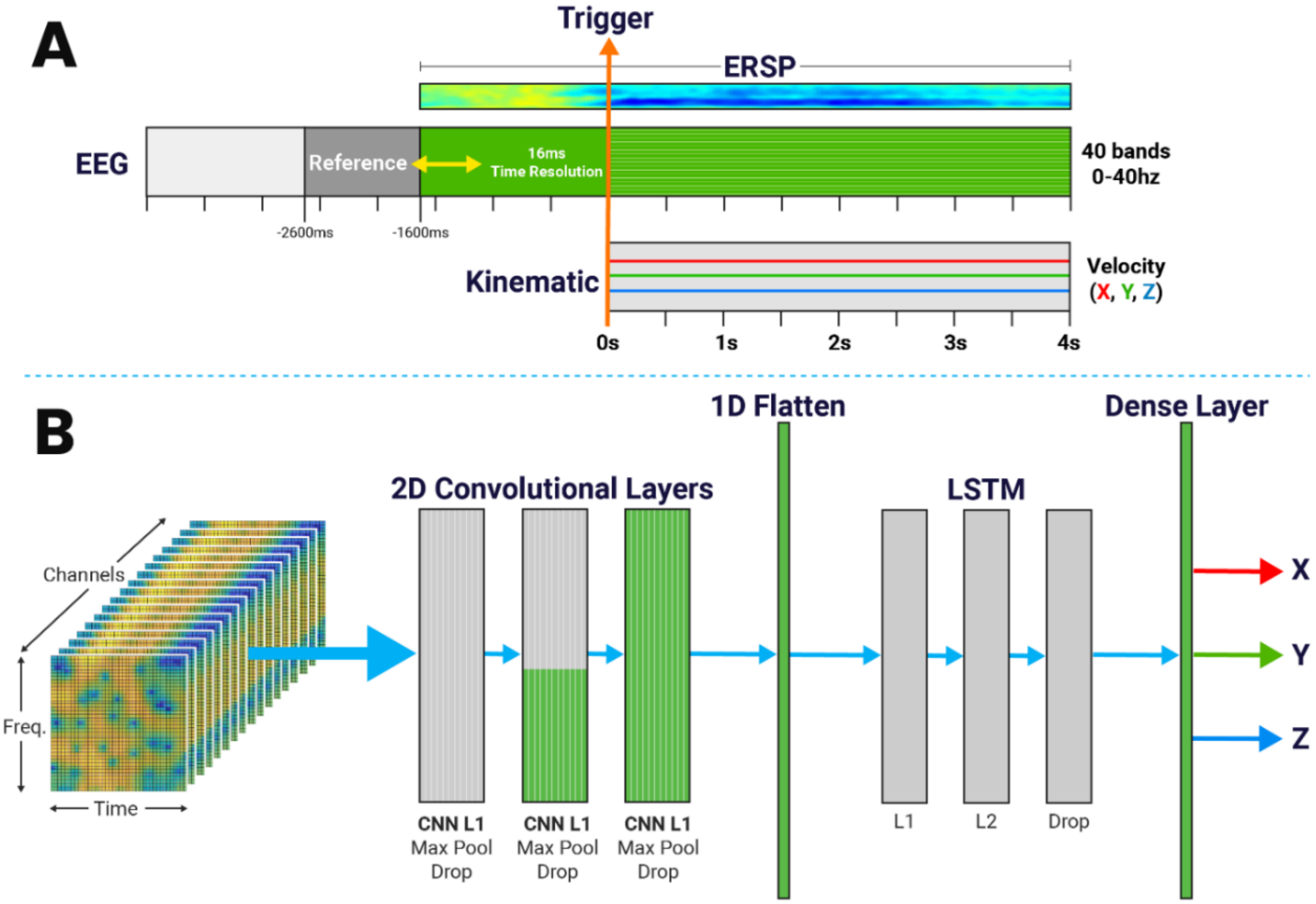

*Figure 3: EEG to ERSP conversion and CNN-LSTM architecture. (A) ERSP images (0–40 Hz, 1 Hz resolution; -640 ms to movement onset, 16 ms steps) capture motor-related spectral changes for trajectory decoding. (B) The CNN-LSTM model integrates convolutional layers for spatial feature extraction and stacked LSTM layers for temporal sequence learning, enabling decoding of continuous 3D motion from EEG signals.*

80% of the data were used for training and 20% for validation. The learning rate was initialised at $1 \times 10^{-3}$ and training proceeded for 12 epochs with early stopping based on validation loss. A batch size of 12 was used for training, and batch = 1 for online decoding to enable zero-lag inference. Each temporal window spanned 96 ms, providing a balance between temporal precision and model efficiency.

Model refinement employed Asynchronous Successive Halving Algorithm (ASHA) via Ray Tune [53], allowing adaptive resource allocation and early termination of under-performing models during training. Two hundred separate hyperparameter configurations explored learning rate, batch size, convolutional depth, filter count, kernel size, dropout, and activation function combinations (Table 1). The best configuration per modality (screen or VR) was selected for online control, with separate networks used to account for feedback-dependent modulation of MI patterns.

*Table 1: CNN-LSTM hyperparameter search space. Ranges and values tested during ASHA-based optimisation in Ray Tune, exploring learning rate, architecture depth, kernel size, dropout, and activation parameters for model refinement across feedback modalities.*

| Parameter | Range / Values Tested | Notes |
|---|---|---|
| **Learning rate** | 1e-6 – 1e-2 (log-uniform) | Convergence tuning |
| **Batch size** | 6, 12, 24, 48 | Efficiency vs. update frequency |
| **Conv. layers** | 2 – 4 | Filter counts = 16–32 |
| **Kernel size** | 3, 5, 7 | Receptive field variation |
| **Dropout** | 0.1 – 0.4 | Pooling + LSTM regularisation |
| **Activation** | ReLU, tanh | Evaluated for CNN / LSTM |
| **Bias regularisation** | 1e-4 – 1e-3 (log-uniform) | Overfitting control |
| **LSTM units / layers** | 1–2 layers; 20–100 units (L1) | Selected by validation performance |

## 2.5 Decoder Evaluation Strategies

DA was assessed via Pearson's correlation coefficient ($r$) between decoded and target velocity trajectories. Correlations were computed per trial over the 2 s movement window (rest to target hold) for each axis (x, y, z), then averaged across trials to yield per-target DA values, capturing both accuracy and inter-trial consistency during imagined movement.

Three complementary evaluation strategies were applied to characterise different aspects of decoder and user performance. Fixed Decoder Generalisation (FDG) reflects the actual online BCI performance recorded during this study. Session 1 trained decoders were deployed without retraining across Sessions 1–5, and aggregate FDG values reported in Table 2 and §3.1 represent means across these five sessions. Therefore, any performance changes across sessions reflect user adaptation and learning rather than algorithmic improvement, and FDG provides a more ecologically relevant approximation of real-world BCI operation, where frequent recalibration may not be feasible. Sequential Adaptive Training (SAT) is an offline evaluation in which decoders were trained on one session's data and tested on the subsequent session (S1→1, S1→2, S2→3, S3→4, S4→5), estimating the combined effects of incremental decoder retraining and user learning. SAT quantifies the potential benefit of periodic recalibration relative to a fixed decoder. Within-Session Reconstruction (WSR) is an offline evaluation that establishes an upper-bound estimate of decoder capacity by training and testing within the same session (S1→1, S2→2, S3→3, S4→4, S5→5). To avoid circular inference, WSR employed five-fold cross-validation within each session: data were partitioned into five non-overlapping folds stratified by target, with the model trained on four folds and tested on the held-out fold, rotating across all five splits. The reported r values are the mean across folds. Because WSR eliminates cross-session variability, any gap between WSR and FDG/SAT can be attributed to generalisation challenges and user-decoder adaptation dynamics rather than to fundamental decoder limitations.

## 3 Results

DA was evaluated across three complementary strategies to distinguish actual online BCI performance from offline estimates of decoder optimisation and capacity. Results are presented in order of ecological validity: FDG (actual online performance), SAT (offline sequential retraining), and WSR (offline within-session upper bound).

### 3.1 Feedback Modality Effects on Decoding Accuracy

VR feedback significantly outperformed screen feedback across all three decoder strategies and movement dimensions (Table 2). Overall mean correlation improvements ranged from 8.9% to 13.0% (all $p \leq 0.002$), with effect sizes from large ($d = 1.42$) to very large ($d = 2.05$). All individual axes demonstrated significant VR superiority across strategies, with the z-axis (forward movement) showing the largest absolute gains ($\Delta r = 0.087–0.170$), suggesting stereoscopic depth cues particularly enhanced forward trajectory decoding. The x-axis (lateral movement) exhibited the lowest absolute correlations in both conditions due to increased error in bi-directional movements but showed substantial relative improvements under VR ($\Delta r = 0.129–0.137$). These patterns were corroborated by linear mixed-effects modelling, which confirmed significant main effects of feedback type and axis direction without a significant feedback-by-axis interaction (Section 3.3).

Under FDG, which reflects actual online BCI performance with no decoder retraining, VR feedback yielded a mean r of 0.668 compared with 0.538 for screen ($\Delta r = 0.130$, $t(9) = 4.48$, $p = 0.002$, $d = 1.42$), indicating that VR elicited inherently more decodable neural patterns even when the decoder could not adapt. SAT, in which decoders were retrained offline on the immediately preceding session's data and tested on the subsequent session, yielded intermediate performance (screen: $r = 0.577$; VR: $r = 0.670$), exceeding FDG and indicating that periodic retraining provides incremental benefit beyond user adaptation alone. As expected, WSR produced the highest correlations (screen: $r = 0.672$; VR: $r = 0.762$), confirming that decoder capacity remained high throughout the protocol when cross-session variability was removed. The substantial gap between FDG and WSR ($\Delta r \approx 0.09–0.13$) quantifies the generalisation cost of deploying a fixed decoder across sessions. VR's relative advantage remained consistent (8.9–13.0%) regardless of strategy, demonstrating robust benefits of VR feedback that are not contingent on any decoder training paradigm.

Individual participant DA values for all three evaluation strategies are provided in Supplementary Tables 1–3; a summary performance dashboard is shown in Supplementary Fig. 1.

*Table 2: Decoding accuracy (r) comparing screen and VR feedback across three decoder strategies. Values represent mean correlation coefficients across participants (n=10) for each movement axis. All VR vs screen comparisons reached significance (p < 0.05). Note: FDG = Fixed Decoder Generalisation; SAT = Sequential Adaptive Training; WSR = Within-Session Reconstruction. Δr represents overall mean difference (VR - screen). Statistical tests are paired t-tests on overall mean values.*

| Strategy | Feedback | X-axis Mean (SD) | Y-axis Mean (SD) | Z-axis Mean (SD) | Overall Mean (SD) | Δr | t(9) | p | d |
|---|---|---|---|---|---|---|---|---|---|
| **FDG** S1 Fixed Decoder | **Screen** | 0.174 (0.075) | 0.765 (0.132) | 0.674 (0.163) | 0.538 (0.098) | — | — | — | — |
| | **VR** | 0.303 (0.085) | 0.856 (0.044) | 0.844 (0.050) | 0.668 (0.044) | +0.130 | 4.48 | 0.002 | 1.42 |
| **SAT** Sequentially Retrained | **Screen** | 0.209 (0.066) | 0.799 (0.059) | 0.722 (0.067) | 0.577 (0.041) | — | — | — | — |
| | **VR** | 0.345 (0.107) | 0.840 (0.037) | 0.826 (0.044) | 0.670 (0.037) | +0.094 | 6.49 | <0.001 | 2.05 |
| **WSR** Same Session Trained Tested | **Screen** | 0.390 (0.073) | 0.840 (0.057) | 0.786 (0.052) | 0.672 (0.045) | — | — | — | — |
| | **VR** | 0.527 (0.080) | 0.884 (0.020) | 0.873 (0.018) | 0.762 (0.030) | +0.089 | 5.83 | <0.001 | 1.84 |

### 3.2 User Adaptation and Learning Dynamics

User adaptation was assessed using three complementary training strategies to distinguish feedback-dependent learning effects from general skill acquisition across sessions. To isolate the influence of feedback modality, feedback-dependent adaptation was examined across Sessions 1–5, where participants alternated between screen and VR feedback conditions on a session-by-session basis (Figure 4A-B). To capture overall learning trajectories independent of feedback type, feedback-independent adaptation was evaluated by combining both modalities across the full 10-session protocol (Figure 4C).

Sessions 1–2 employed 100% trajectory assistance as visually guided motor imagery without decoder-based feedback to establish offline calibration models. From Session 3 onwards, pre-trained decoders were deployed for real-time control with progressively reduced trajectory assistance (65% in S3–4, 60% in S5–6, 50% in S7–8, 40% in S9–10). By maintaining fixed decoders in the FDG approach, performance changes reflect user adaptation rather than decoder retraining, mirroring actual online session performance. SAT and WSR served as complementary analyses to assess the robustness of findings across periodic (session to session) decoder retraining strategies and to establish performance upper bounds, respectively.

#### 3.2.1 Feedback-Dependent Adaptation (Sessions 1-5 : Per Feedback)

VR feedback demonstrated superior and more stable adaptation compared to screen feedback across all training approaches (Figure 4B). Note that S1 values are identical across FDG and WSR strategies, as Session 1 served as both the calibration session and the first within-session reconstruction fold; from S2 onwards the strategies diverge as cross-session generalisation and decoder retraining effects begin to separate. Under FDG, which captures actual online performance, VR maintained consistently high accuracy across sessions (S1: r = 0.771, S2: r = 0.670, S3: r = 0.624, S4: r = 0.627, S5: r = 0.647), showing resilience to cross-session generalisation despite no decoder retraining. Screen feedback, by contrast, showed marked difficulty sustaining performance with a fixed decoder (S1: r = 0.658, S2: r = 0.500, S3: r = 0.517, S4: r = 0.497, S5: r = 0.516), with accuracy dropping sharply after calibration and remaining suppressed throughout online sessions. This divergence indicates that VR elicited neural patterns that were not only more decodable but also more stable across sessions, whereas screen feedback required greater session-specific neural adaptation that a fixed decoder could not accommodate. SAT revealed that sequential decoder retraining partially mitigated the screen condition's generalisation deficit. Under VR, SAT showed a dip at the first online test session before recovering across subsequent sessions (S1: r = 0.771, S2: r = 0.670, S3: r = 0.554, S4: r = 0.690, S5: r = 0.666), with performance recovering to above-FDG levels by S4 following decoder retraining on online session data. Under screen feedback, SAT demonstrated a broadly similar pattern of an initial dip followed by partial recovery across sessions (S1: r = 0.658, S2: r = 0.500, S3: r = 0.546, S4: r = 0.613, S5: r = 0.566), though performance remained consistently below VR levels throughout.

As expected, WSR produced the highest and most stable values in both conditions by eliminating cross-session variability. VR achieved high and largely stable decoder fidelity across sessions (S1: r = 0.771, S2: r = 0.717, S3: r = 0.783, S4: r = 0.759, S5: r = 0.776). Performance was modestly lower at S2 (r = 0.717) relative to the calibration session (r = 0.771), before recovering to its highest value at S3 (r = 0.783) and remaining stable across S4–S5, confirming that VR feedback elicited consistently decodable neural patterns across sessions. In contrast, screen feedback achieved comparable within-session fidelity across sessions (S1: r = 0.658, S2: r = 0.660, S3: r = 0.722, S4: r = 0.681, S5: r = 0.642). Screen WSR showed no equivalent dip at S2, instead remaining stable from calibration through to online sessions, indicating that screen-trained decoders could reach reasonable accuracy when trained and tested within the same session. That this performance did not transfer to the FDG condition reinforces the interpretation that screen feedback produces more session-specific neural representations, limiting cross-session decoder robustness.

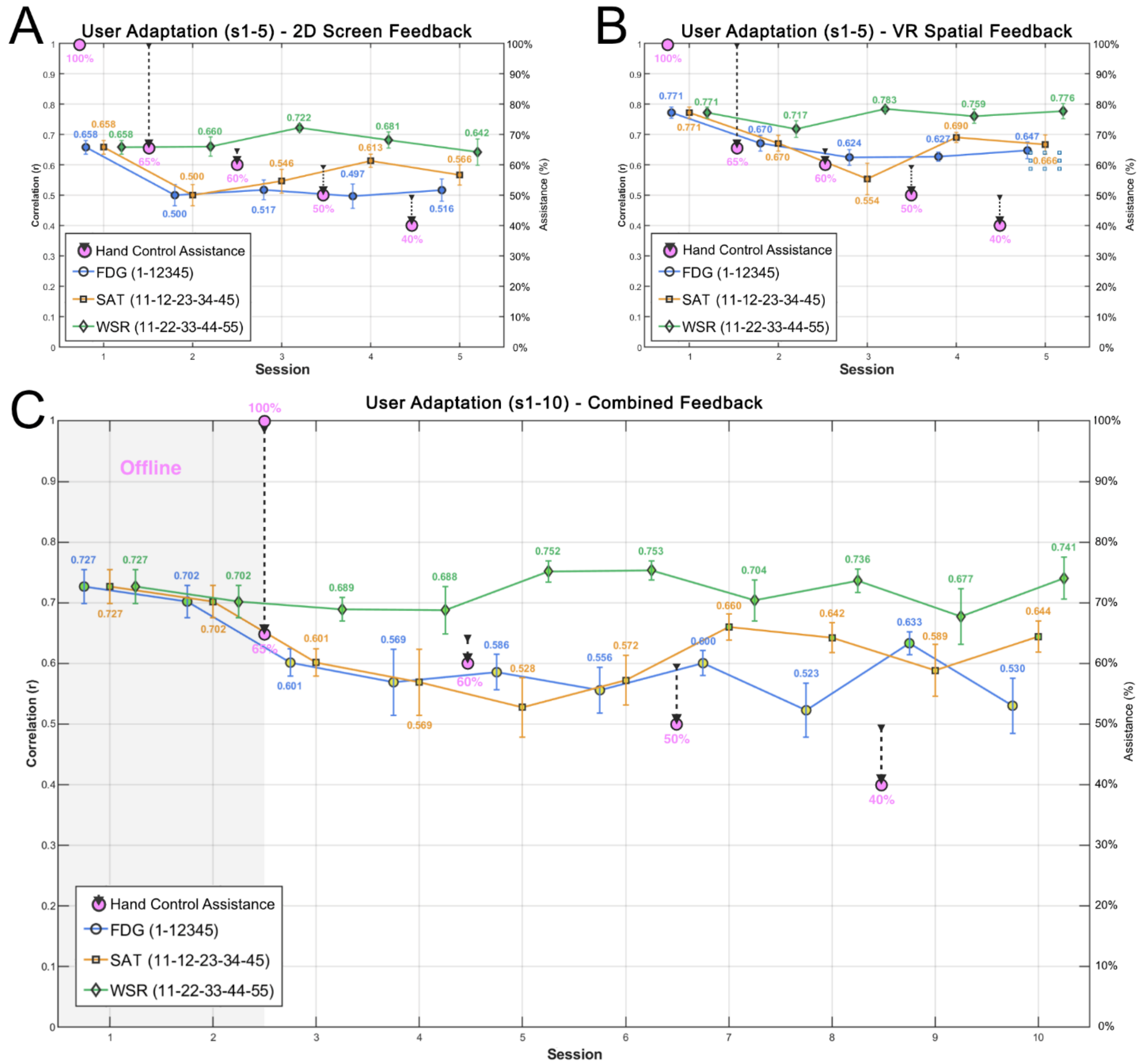

*Figure 4: User adaptation across training strategies and feedback conditions. Pearson's r for Fixed-Decoder Generalisation (FDG, blue), Sequential Adaptive Training (SAT, orange), and Within-Session Reconstruction (WSR, green). Purple markers show trajectory assistance (100% offline to 40% by S9-10). (A) Screen feedback shows poor FDG generalisation across online sessions (S2–5: r = 0.497–0.517). (B) VR feedback maintains stable FDG performance across online sessions (r = 0.624–0.771). (C) Combined 10-session data shows adaptation despite reduced assistance, with WSR (r = 0.677–0.753) establishing the decoder capacity upper bound. n = 10.*

### 3.2.2 Feedback-Independent Adaptation (Sessions 1-10 : Combined Feedback)

When combining both feedback modalities across the full 10-session protocol, distinct adaptation patterns emerged across training strategies (Figure 4C). During offline calibration, all strategies were equivalent, with FDG recording r = 0.727 at S1 and r = 0.702 at S2. The first substantive drop occurred at S3 (r = 0.601), coinciding with the transition to online decoder-based control at 65% assistance, with performance reaching a local minimum at S4 (r = 0.569) before fluctuating across S5–10 (r = 0.586, 0.556, 0.600, 0.523, 0.633, 0.530). The alternating pattern across later sessions reflects the interleaved feedback modality design and progressive assistance reductions. Critically, despite progressively reduced assistance from 65% (S3–4) to 40% (S9–10), participants maintained functional control throughout, demonstrating successful adaptation to decreasing support levels with a fixed decoder. SAT performance was identical to FDG through S1–4, as the sequential retraining protocol had not yet introduced a retrained decoder at this stage (S1: r = 0.727, S2: r = 0.702, S3: r = 0.601, S4: r = 0.569). SAT diverged from S5 onwards, when the first retrained decoder was deployed, producing an initial dip to r = 0.528 before recovering across S6–10 (r = 0.572, 0.660, 0.642, 0.589, 0.644). This recovery pattern is consistent with the within-feedback SAT results in Section 3.2.1, where the first retrained decoder produced a transient dip before stabilising. The improvement over FDG was modest across the online retraining phase (mean r across S5–10: SAT = 0.606 vs FDG = 0.571), indicating that user adaptation rather than decoder updates was the primary driver of sustained performance across sessions.

Consistent with its role as a within-session upper bound, WSR maintained the highest and most stable decoder fidelity across the full 10-session protocol (S1–10: r = 0.677–0.753, mean r = 0.717), with no evidence of degradation across progressive assistance reductions. The substantial gap between WSR and both cross-session strategies (ΔWSR-FDG = 0.089-0.213 across sessions) reflects the cost of deploying fixed decoders against non-stationary neural signals, where gradual shifts in user strategy and neural representation cannot be accommodated without retraining, rather than an inherent ceiling in decoder capacity. The sustained WSR performance throughout the protocol confirms that neural signal quality remained high across all sessions, supporting the interpretation that cross-session performance patterns in FDG and SAT reflect the combined influence of progressive assistance reduction and user-decoder adaptation dynamics rather than fundamental signal degradation.

### 3.3 Statistics and Reproducibility

Pairwise comparisons of feedback conditions within each decoder strategy (Figure 5) were conducted using paired t-tests with Cohen's d as the effect size measure. No correction for multiple comparisons was applied to these pairwise tests, as the LMM constitutes the primary inferential framework; the paired t-tests serve as descriptive summaries of condition differences within each strategy and axis. To provide a more comprehensive inferential framework that accounts for the repeated-measures structure of the data, linear mixed-effects modelling (LMM) was fitted in R [54] using FDG data, with DA as the dependent variable, feedback type (screen, VR; reference: screen) and axis direction (x, y, z) as fixed effects, and participant-specific random intercepts and random slopes for feedback. For the additive models, axis was first coded with x as the reference level and then relevelled with y as the reference level to obtain the y-z comparison. The interaction model used x as the reference level. Models were estimated using restricted maximum likelihood (REML) for reporting fixed-effect parameter estimates, and p-values for fixed effects were obtained using Satterthwaite's approximation to the denominator degrees of freedom as implemented in the lmerTest package [55] in R [54]. Maximum likelihood (ML) was used for model comparisons and calculations of effect size. To avoid singular boundary fits in reduced ML models, the ML models used for effect size estimation were fit with an uncorrelated random-effects structure. Cohen's $f^2$ for each fixed effect was computed as the change in marginal $R^2$ when that predictor was excluded from the full model, with the resulting reduced model compared back to the full model to quantify the unique variance explained by each predictor.

The results revealed a significant main effect of feedback type ($f^2 = 0.39$) and axis direction ($f^2 = 3.89$), without a significant feedback-by-axis interaction (VR × y-axis: $\beta = -0.117$, $p = 0.277$; VR × z-axis: $\beta = 0.153$, $p = 0.155$), consistent with the relatively uniform VR advantage across axes observed in Table 2. By conventional benchmarks (Cohen, 1988 [56]), both represent large effects ($f^2 \geq 0.35$), with the axis direction effect reflecting the substantial performance disparity between horizontal and vertical/forward movements. VR feedback demonstrated a stable advantage across all axis directions ($\beta = 0.43$, $SE = 0.1$, $t(292) = 4.45$, $p = 1.25 \times 10^{-5}$), with movements along the y- and z-axes producing substantially greater DA than movements along the x-axis (y vs. x: $\beta = 1.93$, $SE = 0.05$, $t(292) = 35.6$, $p = 2.87 \times 10^{-108}$; z vs. x: $\beta = 1.75$, $SE = 0.05$, $t(292) = 32.3$, $p = 2.1 \times 10^{-98}$). Additionally, in the relevelled additive model (with y as the axis reference) performance along the y-axis was modestly but significantly greater than along the z-axis (y vs. z: $\beta = 0.18$, $SE = 0.05$, $t(292) = 3.31$, $p = 1.06 \times 10^{-3}$). In the interaction model, inclusion of random effects increased explained variance from 80.8% to 85.8%, indicating a strong influence of fixed effects on the model whilst between-participant variability remained modest. VR feedback produced a consistent boost in performance across all axis directions, with no evidence that this advantage depended on movement direction.

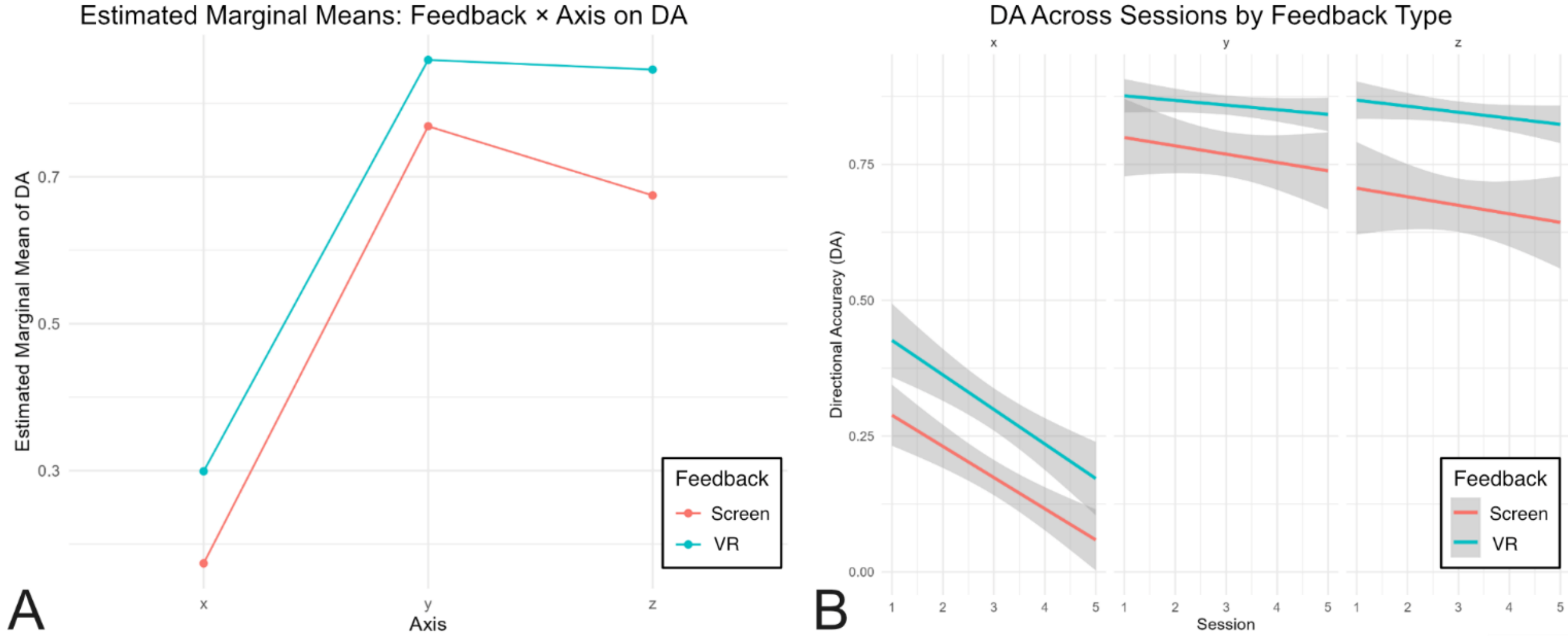

*Figure 5: (A) Estimated marginal means of decoding accuracy (DA) for each axis direction under screen and VR feedback conditions, derived from the additive linear mixed-effects model. (B) Decoding accuracy (DA) across sessions by feedback type, shown separately for x (horizontal), y (vertical), and z (forwards) movements. Solid lines represent model-estimated trends; shaded bands indicate 95% confidence intervals. Session-related slope coefficients from the LMM indicate a modest decline in DA across sessions for both feedback conditions: x-axis (screen: -0.057, VR: -0.064), y-axis (screen: -0.015, VR: -0.009), and z-axis (screen: -0.016, VR: -0.011). Negative values indicate performance decline across sessions.*

To examine whether DA changed significantly across sessions, session number was included as a continuous predictor in a separate LMM with feedback type, axis direction, and their interactions as additional fixed effects, and participant as a random intercept. Fixed effects were evaluated using t-tests with Satterthwaite-approximated degrees of freedom. DA declined significantly across sessions, with no evidence that the rate of decline differed by feedback type, and therefore, no indication of a positive learning effect across sessions for either feedback condition (Figure 5). The decline was most pronounced along the x-axis (screen: -0.057; VR: -0.064 per session) and minimal along the y- and z-axes (screen: -0.015 to -0.016; VR: -0.009 to −0.011), where baseline performance was highest and neural representations were most spatially distinct. There was no significant feedback × session interaction, indicating that VR feedback, while producing consistently higher DA, did not differentially protect against this session-related decline.

### 3.4 Topographical and Functional Connectivity

Grand-average scalp maps (N = 10) were computed for six frequency bands: delta (0–4 Hz), theta (4–8 Hz), alpha (8–12 Hz), low beta (12–18 Hz), high beta (18–28 Hz), and gamma (28–40 Hz). Power changes are shown as movement minus rest for screen and VR, with a VR>screen difference row to highlight condition effects (Figure 6).

Across conditions, alpha and beta bands exhibited the expected motor-related desynchronisation over central regions. Compared with screen, VR produced stronger and more focal alpha/beta suppression over bilateral sensorimotor and posterior parietal sites, consistent with enhanced engagement of motor–visuospatial systems during embodied control. The VR/screen difference maps show positive differences centred over central–parietal areas in alpha and extending fronto-centrally in high beta, indicating larger movement-related modulation in VR. In delta and theta, screen showed broader posterior patterns, whereas VR displayed relatively stronger frontal and midline modulation, aligning with increased preparatory and control demands under immersive feedback. In gamma, the VR condition showed more extensive fronto-central and parietal modulation than screen, with the VR/screen difference map indicating widespread but especially central–parietal increases.

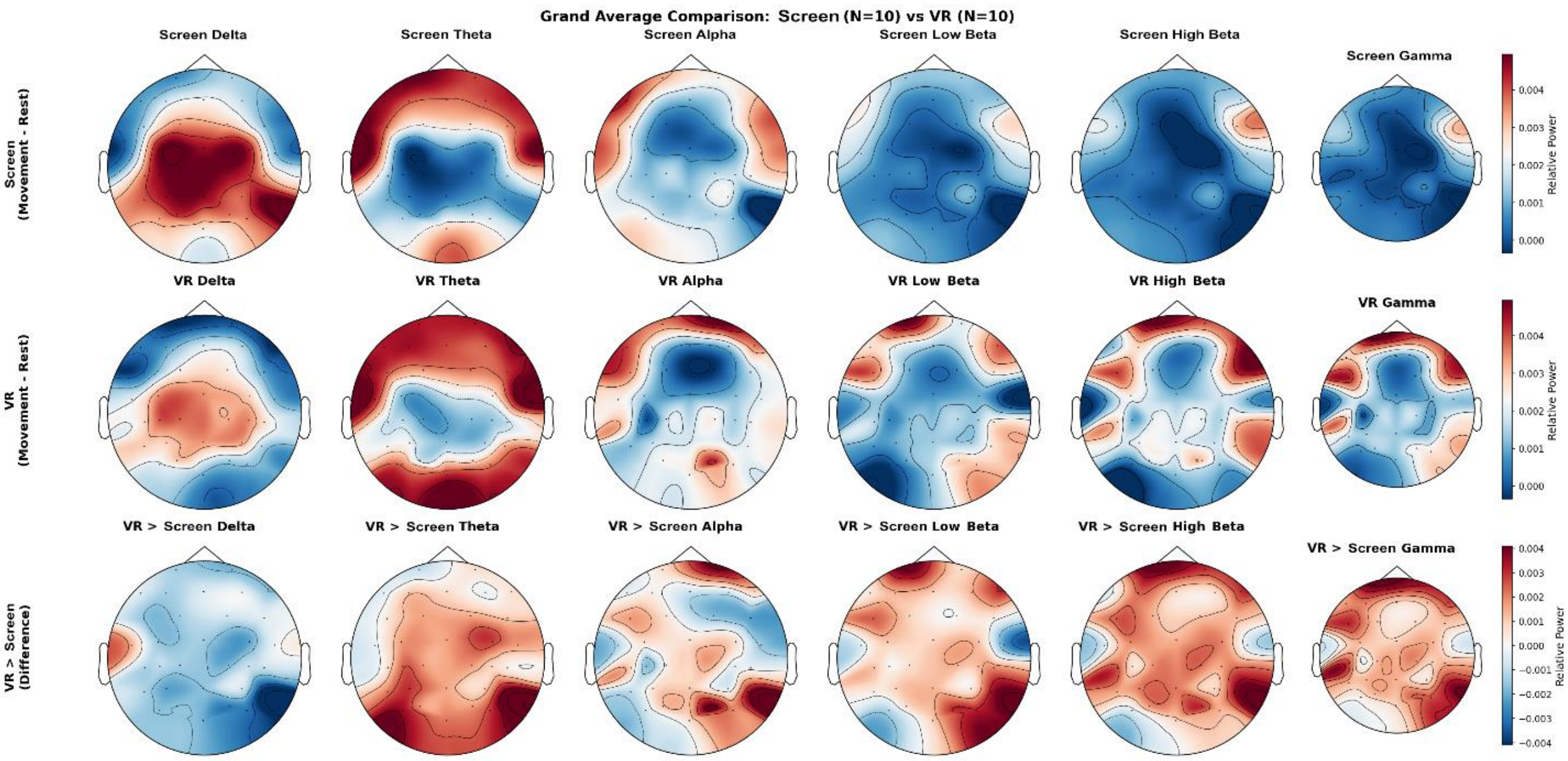

*Figure 6: Topographical analysis of movement-related power. Grand-average ERD/ERS maps (N = 10) across six frequency bands showing movement-rest power for screen and VR feedback, with VR > screen differences highlighting stronger, more focal alpha/beta desynchronisation.*

Task-related FC was assessed using the envelope of imaginary coherence (EIC), a volume-conduction-robust measure of amplitude-coupled phase-lagged synchrony that is insensitive to spurious zero-phase correlations arising from field spread [57,58]. EIC was computed for each electrode pair from the cross-spectral density matrix of the EEG signal. Source-level EIC differences between the reaching period and a resting baseline were computed using an inverse solution applied to the 32-channel montage, with source regions defined by a pre-specified visuomotor template [18]. A 1.5 s epoch from the reaching period was compared against a 1.5 s baseline epoch across both feedback conditions (Figure 7A). Connectivity patterns were examined across six frequency bands: delta (0–4 Hz), theta (4–8 Hz), alpha (8–12 Hz), low beta (12–18 Hz), high beta (18–28 Hz), and gamma (28–40 Hz). Statistically significant task-related connections were identified using non-parametric Monte Carlo permutation testing (N = 1000 permutations, $p < 0.05$), in which trial labels were randomly reassigned to generate a null distribution of maximum EIC statistics; connections exceeding the 95th percentile of this distribution were retained as statistically significant. For visual clarity, a secondary display threshold (EIC difference > 0.55) was subsequently applied to highlight the strongest task-related connections within each condition (Figure 7B); this threshold was selected empirically to retain only connections in the upper quartile of observed EIC values while preserving interpretable network structure. All connections shown passed the permutation test. Additional above-threshold connections include superior parietal lobule coupling under VR across four frequency bands (delta: aINS-SPL = 0.566; theta: SPL-SFG = 0.573; alpha: SMG-SPL = 0.570, PCC-SPL = 0.565; gamma: aINS-SPL = 0.559), and dorsal anterior cingulate cortex coupling under VR in the delta (dACC-IFG = 0.574) and alpha (aINS-dACC = 0.568) bands.

The overall network architecture differed substantially between conditions across frequency bands (Table 3). The screen condition was characterised by stronger long-range posterior-anterior connectivity, with prominent coupling between temporal, parietal, and frontal regions across delta, theta, and low-beta bands (e.g., MTG-OL: 0.59, PCC-SFG: 0.59, SFG-SPL: 0.58). In contrast, VR feedback consistently engaged motor cortex in strongly coupled networks across five of the six bands examined, a pattern largely absent in the screen condition, which showed motor cortex coupling in only two bands above threshold. VR feedback also produced more extensive connectivity involving the anterior insula across all frequency bands (delta through gamma), whereas screen feedback showed aINS coupling confined primarily to the delta and alpha bands. The posterior cingulate cortex similarly showed a condition-dependent shift: under screen feedback, PCC coupling was limited to frontal and occipital regions (theta, low-beta), whereas under VR it extended to motor and executive areas across theta, alpha, low-beta, high-beta, and gamma bands. These exploratory findings are interpreted with caution given the source localisation constraints of the 32-channel montage; they are presented as supporting evidence for the primary decoding results rather than definitive neuroanatomical claims, and the specific anatomical regions identified should be treated as approximate.

*Table 3: Task-related source-level functional connectivity (EIC difference, task vs. baseline) for screen and VR feedback across six frequency bands. The two strongest connections per condition per band are shown (EIC difference > 0.55; all connections significant via Monte Carlo permutation testing, N = 1000, p < 0.05); additional above-threshold connections are discussed in the text. Findings are exploratory given the 32-channel montage. EIC: envelope of imaginary coherence. Abbreviations: SMG, supramarginal gyrus; aINS, anterior insula; MFG, middle frontal gyrus; MTG, middle temporal gyrus; FP, frontal pole; dACC, dorsal anterior cingulate cortex; PCC, posterior cingulate cortex; SFG, superior frontal gyrus; IFG, inferior frontal gyrus; MC(L/A), motor cortex (lateral/anterior); SPL, superior parietal lobule; OL, occipital lobe.*

| | **Screen Feedback** | | **VR Feedback** | |
|---|---|---|---|---|
| **Frequency Band** | **Strongest connections** | **EIC** | **Strongest connections** | **EIC** |
| Delta (0–4 Hz) | MTG - OL | 0.59 | MTG - SFG | 0.59 |
| | SFG - OL | 0.58 | MFG - FP | 0.58 |
| Theta (4–8 Hz) | MC(A) - SFG | 0.58 | aINS - MC(A) | 0.59 |
| | SPL - MC(A) | 0.57 | aINS - SFG | 0.58 |
| Alpha (8–12 Hz) | FP - aINS | 0.56 | MC(L) - IFG | 0.58 |
| | aINS - SMG | 0.56 | OL - IFG | 0.58 |
| Low Beta (12–18 Hz) | SMG - SPL | 0.59 | PCC - MC(A) | 0.59 |
| | PCC - SFG | 0.59 | PCC - IFG | 0.57 |
| High Beta (18–28 Hz) | SFG - SPL | 0.58 | MTG - aINS | 0.57 |
| | FP - SPL | 0.58 | PCC - IFG | 0.57 |
| Gamma (28–40 Hz) | FP - SFG | 0.59 | FP - MC(L) | 0.58 |
| | FP - MFG | 0.56 | PCC - MFG | 0.58 |

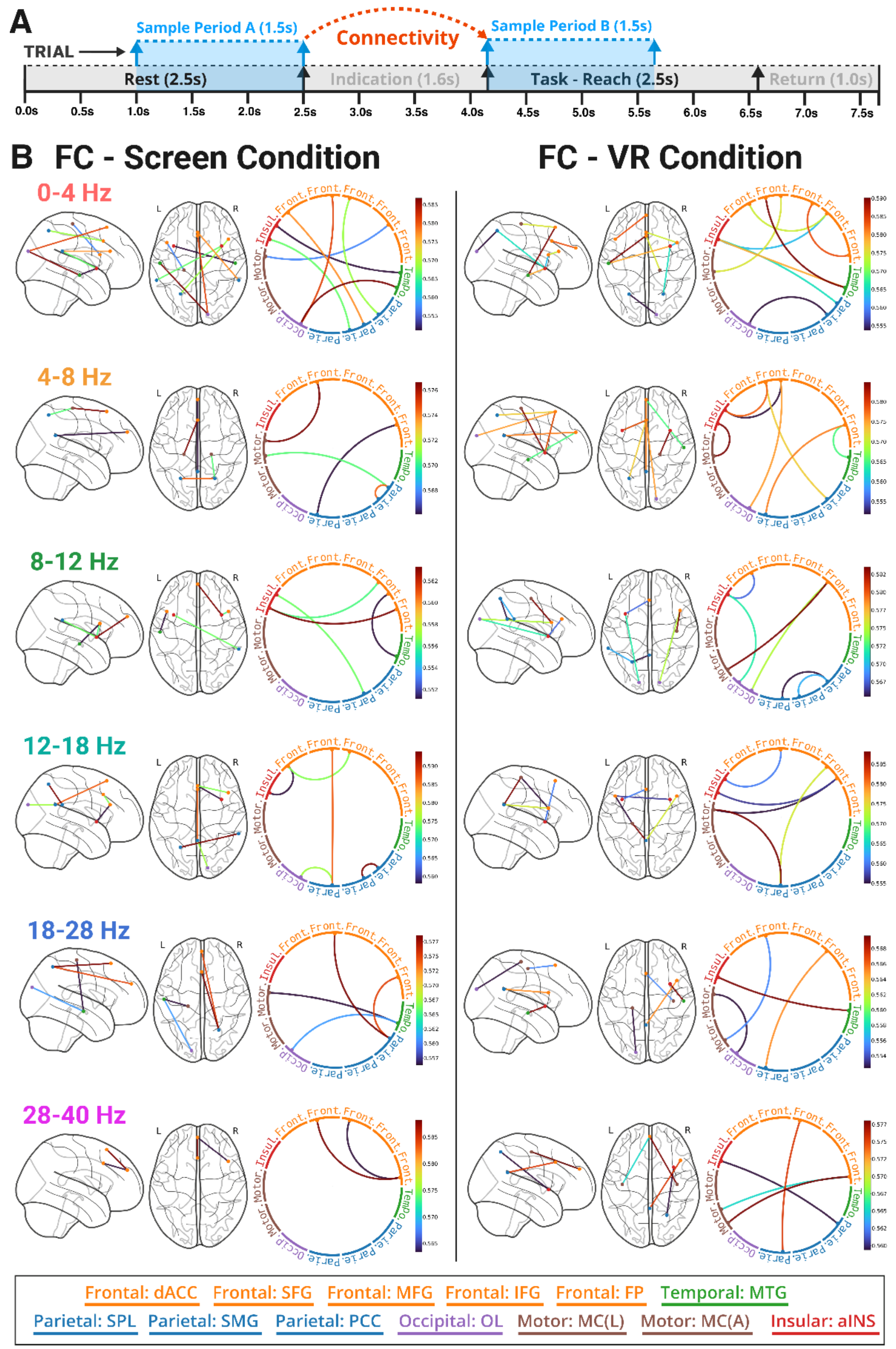

*Figure 7: (A) FC was computed from 1.5 s EEG segments during the reaching period and compared with 1.5 s resting-period segments across both feedback conditions. (B) Connectivity maps (relative strength > 0.55) compare screen (left) and VR (right) feedback over six frequency bands (0–40 Hz). Colour-coded lines indicate connection strength between brain regions.*

# 4 Discussion

This study investigated how embodied spatial feedback delivered through VR influences decoding accuracy, learning dynamics and cortical network reorganisation in a real-time, continuous 3D motor imagery decoding for 3D virtual limb control. Ten participants completed ten sessions of executed and imagined reaching movements decoded by a CNN-LSTM architecture, with performance evaluated across three complementary training strategies (FDG, SAT, WSR) and two feedback modalities (screen, VR). The principal findings demonstrate that VR feedback significantly and consistently enhances DA across all evaluation strategies and movement dimensions, supports maintained functional control despite progressively reduced trajectory assistance albeit without evidence of progressive improvement in DA across sessions, produces more focal sensorimotor-parietal desynchronisation in topographical maps and drives enhanced motor-frontal functional connectivity.

## 4.1 Decoding Accuracy and Feedback Dynamics

VR feedback significantly outperformed screen feedback across all three decoder strategies and movement dimensions, with mean correlation improvements of 8.9–13.0% (all $p \leq 0.002$, $d = 1.42–2.05$) that remained consistent regardless of strategy, demonstrating that the benefits of embodied spatial feedback are robust and not an artefact of any specific decoder or optimisation approach.

The z-axis (forward movement) showed the largest absolute DA gains ($\Delta r = 0.087–0.170$), suggesting that stereoscopic depth cues and egocentric spatial encoding particularly benefit movements relying on depth perception. The x-axis (lateral) exhibited the lowest correlations in both conditions due to bi-directional velocity overlap [11,43], yet showed the largest relative improvements under VR ($\Delta r = 0.129–0.137$), indicating that VR feedback partially mitigates this inherent decoding challenge. The strategy hierarchy (WSR > SAT > FDG) held in both conditions, confirming that decoder retraining provides incremental benefit, but that VR feedback enhances performance independently of decoder optimisation schedule.

These findings extend prior demonstrations that VR feedback improves discrete MI-BCI classification to continuous trajectory decoding, where the mapping between neural activity and kinematics is graded rather than categorical [39–41]. The very large effect sizes, particularly for SAT ($d = 2.05$), exceed those typically reported in discrete MI-BCI comparisons [39,41], suggesting that the benefit of spatial alignment may be amplified in continuous paradigms where ongoing visuomotor correspondence provides sustained rather than momentary performance cues. The maintenance of VR superiority under FDG ($p = 0.002$), where a fixed decoder was applied without retraining, carries practical significance: it indicates that VR feedback elicits inherently more decodable and generalisable neural patterns, reducing dependence on frequent recalibration for real-world BCI deployment [42]. This aligns with Wolpaw's (2025) synthetic heksor framework, which holds that improved signal analysis raises mean BCI accuracy without reducing session-to-session variability because the core bottleneck is the user-decoder adaptive relationship rather than the decoder alone [28]. The persistent VR advantage under fixed decoders suggests that embodied spatial feedback strengthens the neural component of this relationship, driving motor imagery representations closer to the action-like patterns that yield inherently more stable and decodable signals across sessions.

The CNN-LSTM decoder architecture itself contributed to the high absolute DA values observed across conditions. The ERSP-CNN-LSTM, optimised via ASHA-based hyperparameter search [18,52], achieved within-session imagined movement correlations (WSR mean $r = 0.762$ under VR; $r = 0.672$ under screen), comparing favourably with established MTD approaches involving physical movement. Pancholi et al. (2022) reported a maximum $r = 0.67$ using source-localised CNN-LSTM decoding of executed hand trajectories [15], while Li et al. (2024) achieved $r = 0.511$ with feature-selected nonlinear regression for continuous upper limb trajectory decoding from imagined movements [23]. That the present within-session values for imagined movement approach or exceed benchmarks obtained from executed movement decoding is noteworthy, given that imagined movements are generally considered harder to decode due to the absence of peripheral motor output. The present results extend these benchmarks to 3D motor imagery of reaching under real-time feedback, suggesting that the combination of ERSP-based time-frequency representation with CNN-LSTM temporal feature extraction provides an effective feature-decoder pipeline for MTD. The capacity of the convolutional layers to extract nonlinear spatial-temporal dependencies from EEG likely contributed to the strong axis-direction effects observed in the LMM ($f^2 = 3.89$), as linear approaches may underrepresent the separable spectral signatures that distinguish vertical and forward movements from lateral movements [11,16]. Korik et al. (2025) similarly demonstrated the effectiveness of LSTM-based architectures for continuous velocity decoding from ERSP features in a lower-limb stepping paradigm [18], and the present findings confirm that this approach generalises to 3D upper-limb motor imagery for virtual limb trajectory control under both feedback modalities.

### 4.2 Training Strategies and User Adaptation

Beyond the feedback modality effects, the three training strategies revealed distinct adaptation dynamics that illuminate where the performance bottleneck lies in cross-session MTD-BCI use. SAT yielded only modest improvements over FDG (mean S5–10: SAT r = 0.606 vs. FDG r = 0.571), indicating that user adaptation rather than decoder updates was the primary driver of sustained performance. The WSR upper bound (mean $r = 0.717$) confirmed that decoder capacity remained high throughout, and that the FDG–WSR gap ($\Delta r \approx 0.089$–$0.213$) likely reflects cross-session generalisation challenges rather than progressive decoder degradation. The ASHA-based hyperparameter optimisation [53] may have contributed to this robust decoder ceiling by producing models that are well-regularised and less prone to overfitting session-specific features, though this also means the remaining performance gap is attributable to neural variability rather than model limitations. This dissociation suggests that future gains are more likely to come from improving cross-session neural consistency than from decoder architecture alone or training decoders on accumulated data across multiple sessions and testing on later sessions. However, the LMM analysis of session effects qualifies this interpretation. Rather than revealing progressive improvement, DA declined modestly but significantly across sessions regardless of feedback modality. This decline was most pronounced along the x-axis (screen: -0.057; VR: -0.064 per session) and minimal along the y- and z-axes (-0.009 to -0.016), where baseline performance was highest and neural representations were most spatially distinct. Crucially, this pattern does not indicate deteriorating signal quality or decoder capacity, as WSR confirmed consistently high within-session fidelity throughout the protocol (mean $r = 0.717$). Instead, the decline likely reflects the compounding effect of progressive assistance reduction from 100% to 40% and the constraints of fixed decoders that cannot accommodate gradual shifts in user strategy. The absence of feedback × session interaction further indicates that VR feedback, while producing consistently higher DA, did not differentially protect against this session-related decline. Wolpaw (2025) reaches a similar conclusion from theoretical grounds, arguing that BCI reliability requires integration of the BCI skill into the negotiated equilibrium maintained by existing motor networks, and that relatively stable decoders best support user learning by providing a consistent adaptive target for the central nervous system [28]. The graduated assistance reduction protocol employed here may have further facilitated this process by progressively transferring control demands to the user, encouraging the development of robust internal motor imagery strategies.

Despite the overall decline in DA, participants maintained functional control throughout the full 10-session protocol as trajectory assistance was progressively reduced from 100% (offline) to 40% (S9–10). When both feedback modalities were combined, the decline from offline calibration into online control (S1: r = 0.727, S3: r = 0.601, S4: r = 0.569) was followed by fluctuation across S5–10 (r = 0.523–0.644), with the LMM confirming that the overall trajectory remained negative rather than reflecting genuine recovery. This distinction is important: while the pattern superficially resembles the typical learning curve reported in BCI training [46], the declining DA indicates that participants did not progressively improve their motor imagery strategies in the manner observed in discrete MI-BCI protocols [59]. Stefano Filho et al. (2022) reported that repeated MI practice with feedback progressively reshapes FC [31], and the connectivity reorganisation observed here is consistent with ongoing neural adaptation, even if that adaptation was insufficient to offset the increasing task demands imposed by assistance withdrawal. These findings suggest that graduated assistance reduction remains a viable scaffolding mechanism for 3D motor imagery BCIs, but that sustaining or improving absolute DA across extended protocols will likely require co-adaptive decoding frameworks capable of tracking gradual shifts in neural strategy [28]. The steeper x-axis decline (approximately three to four times that of the y- and z-axes) further implies that lateral movements, already the most challenging to decode due to bi-directional velocity overlap, are also the most vulnerable to the compounding effects of reduced support and decoder drift, highlighting them as a priority target for direction-specific adaptive training.

### 4.3 Neural Mechanisms: Topographical and Connectivity Evidence

This study extends prior work showing that MI task performance reorganises sensorimotor and visual FC networks relative to resting-state activity [32]. Topographical mapping confirmed the expected alpha/beta desynchronisation over sensorimotor regions, with VR feedback producing stronger and more focal ERD over bilateral sensorimotor and posterior-parietal areas compared to the broader, less spatially specific activation observed under screen feedback. This extends the spectral spatial patterns reported by Korik et al. (2018), who showed that mu, beta, and low-gamma oscillations encode imagined 3D hand kinematics with characteristic central-parietal topographies [8]. Under real-time feedback conditions, similar frequency-specific modulations were observed but were more focal and fronto-parietally distributed, suggesting that VR feedback amplifies and spatially sharpens these cortical dynamics.

The FC analysis supported and extended this interpretation, revealing distinct network organisation between screen and VR feedback conditions. These findings should be interpreted as exploratory given the source localisation constraints of a 32-channel montage and are presented as supporting evidence for the primary decoding results rather than definitive neuroanatomical claims. In the delta band (0–4 Hz), the screen condition showed strong frontal-temporal-parietal coupling (e.g., MTG-OL: 0.59, SFG-OL: 0.58), consistent with reliance on long-range top-down visual processing to interpret abstract feedback. In contrast, VR produced stronger temporal-frontal integration (MTG-SFG: 0.59, MFG-FP: 0.58) alongside direct motor-frontal coupling (SFG-MC(A): 0.57), implicating both prefrontal and motor cortical engagement in spatial context processing under immersive conditions. In the theta band (4–8 Hz), screen feedback showed motor-frontal coupling (MC(A)-SFG: 0.58, SPL-MC(A): 0.57) consistent with top-down motor regulation, whereas VR produced stronger connectivity between the anterior insula and motor-executive regions (Ains-MC(A): 0.59, aINS-SFG: 0.58), alongside cingulate-prefrontal integration (PCC-FP: 0.58), potentially reflecting enhanced interoceptive and embodied processing coordinated with executive monitoring [34,35]. In the alpha band (8–12 Hz), the screen condition showed default-mode and visual-association connectivity (FP-aINS: 0.56, aINS-SMG: 0.56), while VR produced enhanced sensorimotor-executive coupling (MC(L)-IFG: 0.58, OL-IFG: 0.58) alongside aINS–dACC coupling (0.57), linking interoceptive awareness with conflict-monitoring circuitry and consistent with Leeuwis et al. (2021), who reported that alpha-band coupling between motor and visuospatial regions distinguishes high-performing MI-BCI users [30]. In the low-beta band (12–18 Hz), screen feedback was characterised by parieto-cingulate connectivity, with SMG-SPL coupling (0.59) representing the strongest single connection observed across the entire screen dataset, alongside PCC-SFG (0.59), whereas VR promoted direct motor-executive integration (PCC-MC(A): 0.59, PCC-IFG: 0.57), indicating more efficient sensorimotor processing under immersive feedback. In the high-beta band (18–28 Hz), screen feedback showed fronto-parietal coupling (SFG-SPL: 0.58, FP-SPL: 0.58), while VR produced temporal-insular and parieto-frontal connectivity (MTG-aINS: 0.57, PCC-IFG: 0.57). In the gamma band (28–40 Hz), screen feedback showed predominantly prefrontal network coupling (FP-SFG: 0.59, FP-MFG: 0.56), whereas VR revealed executive-motor synergy (FP-MC(L): 0.58, PCC-MFG: 0.58), further demonstrating VR's capacity to engage integrated motor-planning circuits. Across bands, the anterior insula demonstrated more extensive connectivity under VR, particularly in the theta and alpha bands, consistent with its role in interoceptive and embodied processing [34,35], while the posterior cingulate cortex showed stronger motor-executive coupling under VR in the low-beta and high-beta bands, suggesting greater integration between default-mode and sensorimotor networks during immersive feedback. These hub-level reorganisations are consistent with reports that multimodal feedback strengthens integration between motor and higher-order cognitive areas [29], and demonstrate that even within a single sensory modality, the spatial properties of VR feedback can substantially reshape cortical network organisation.

The anterior insula emerged as a prominent hub under VR feedback, showing connections above threshold across all six frequency bands, from delta through gamma, compared with connectivity confined to the delta and alpha bands under screen feedback. This pervasive aINS engagement under VR is consistent with its role in interoceptive awareness and embodied self-representation [34,35], and suggests that spatially congruent, first-person feedback recruits body-referencing neural systems more broadly than abstract 2D display. The dorsal anterior cingulate cortex showed its strongest above-threshold coupling under VR in the delta band (dACC-IFG = 0.574) and alpha band (aINS-dACC = 0.568), reflecting engagement of conflict monitoring and adaptive control circuitry during immersive feedback.

The parietal engagement observed in both topographical and connectivity analyses is particularly relevant to the neural demands of 3D motor imagery for reaching. The superior parietal lobule is a key node in the visuomotor transformation network, integrating target location in egocentric space with motor planning commands during both real and imagined reaching movements [26]. Under VR feedback, SPL connectivity exceeded the display threshold across four frequency bands, delta (aINS-SPL: 0.566), theta (SPL-SFG: 0.573), alpha (SMG-SPL: 0.570, PCC-SPL: 0.565), and gamma (aINS-SPL: 0.559), whereas SPL coupling under screen feedback was largely confined to the theta and high-beta bands. This pattern suggests that VR's spatially congruent, egocentric feedback engages the brain's endogenous visuomotor circuitry more effectively than abstract 2D display, where the spatial correspondence between visual input and motor output is disrupted. Notably, the experimental design accommodated natural visuomotor coordination: participants shifted gaze to the cued target during the indication phase, mirroring the look-then-reach behaviour that normally precedes voluntary arm movements [48]. This ecological alignment between task design and natural gaze-motor coordination may have further recruited SPL-mediated spatial processing during VR trials and may partly explain why VR's decoding advantage was most pronounced along the depth axis, where egocentric spatial encoding demands are greatest. This engagement of visuomotor circuitry under VR is consistent with the argument that BCI skills grounded in the neural patterns associated with the equivalent muscle-based action may produce more stable and compatible motor representations [28].

### 4.4 Limitations and Future Directions

This study provides the first comprehensive evidence that VR feedback enhances DA, learning stability, and neural network organisation in MTD-BCIs. By integrating optimised 3D task design, three distinct decoder training strategies, longitudinal training across ten sessions, and hyperparameter-tuned CNN-LSTM models, it extends prior discrete MI-BCI findings into the domain of real-time continuous 3D control.

Several limitations should be acknowledged. The sample size ($N = 10$), while sufficient for the within-subject longitudinal design, limits statistical power for detecting subtle inter-individual effects and restricts generalisability. The decoder models were intentionally held fixed across sessions within the FDG condition to isolate the contribution of user-driven adaptation from algorithmic improvement. While this design choice enabled a controlled comparison of training strategies, and the WSR condition confirmed that decoder capacity itself remained high throughout (mean $r = 0.717$), it also constrained long-term performance gains, as models could not accommodate evolving neural strategies. The modest improvement of SAT over FDG (mean S5–10: r = 0.606 vs. 0.571) further suggests that periodic single-session retraining provides benefits, but the optimum frequency and scheduling of recalibration remain to be determined alongside mechanisms for real-time co-adaptation. Future work should explore adaptive and transfer-learning frameworks that support mutual adjustment between user and decoder across sessions and examine whether population-level or hybrid models can reduce calibration burden while maintaining individual specificity. The absence of a progressive learning effect across sessions, despite ten sessions of practice (five for each modality), represents a further limitation with respect to claims about long-term skill acquisition. While participants maintained functional trajectory control as assistance was reduced, the LMM-confirmed decline in DA suggests that the graduated assistance protocol, combined with fixed decoders, was insufficient to drive the progressive improvement in motor imagery strategies observed in some discrete MI-BCI training studies [59]. Whether co-adaptive decoding, longer training durations, or modified assistance schedules could elicit genuine progressive improvement in MTD-BCIs remains an open question.

The CNN-LSTM architecture required a full calibration session and offline hyperparameter optimisation via ASHA, precluding same-day decoder updates. Lightweight architectures or incremental learning approaches that support within-session recalibration would improve practical deployability. Finally, the persistent x-axis decoding asymmetry, where lateral bi-directional movements yielded lower and more variable DA than forward

(y) or vertical (z) movements, highlights an unresolved challenge in achieving full 3D control. This functional asymmetry likely reflects overlapping neural representations for lateralised movements and warrants investigation through refined target configurations or direction-specific training protocols.

The decoder targeted Cartesian endpoint velocity derived from wrist position, representing an inverse kinematics solution rather than a joint-angle decomposition of the reaching movement. Whilst this approach is well suited to continuous 3D trajectory control and is consistent with the majority of EEG-based MTD studies [8,12,26], it does not capture the underlying joint-space kinematics of the arm. Decoding joint angles directly would require disentangling the contributions of multiple degrees of freedom at the shoulder, elbow, and wrist, which remains challenging given EEG's spatial resolution and the channel count used for online decoding. Future work targeting joint-level reconstruction, potentially through source-informed features or higher-density recordings combined with simultaneous capture of full arm segment kinematics, could extend the approach towards more biomechanically complete limb control.

Regarding feedback, this study focused exclusively on visual feedback delivered via screen or VR. Multimodal extensions incorporating tactile, proprioceptive, or auditory feedback may further clarify the mechanisms through which embodiment enhances neural reorganisation and could reduce the cognitive load associated with sustained VR use. The FC analysis revealed feedback-dependent network reorganisation across all six frequency bands examined, with condition-specific differences in connectivity pattern rather than band-selective engagement, consistent with prior evidence that MI tasks broadly reorganise sensorimotor and visuospatial networks relative to rest [30–32]. Longitudinal FC analysis tracking connectivity changes across sessions would provide further insight into the training-induced plasticity that the present cross-sectional approach cannot resolve.

The FC analysis employed EIC, which is robust to volume conduction but carries interpretive constraints: EIC detects non-zero phase-lagged amplitude-coupled synchrony but cannot resolve temporal directionality, as signals with substantially different conduction delays may yield equivalent values [57]. Source localisation was performed with a 32-channel montage using a pre-specified visuomotor template, which limits spatial resolution relative to high-density arrays and restricts localisation to template-defined regions. Accordingly, the FC findings are presented as exploratory supporting evidence and the specific anatomical regions identified should be treated as approximate. Future work employing high-density EEG combined with beamformer-based source reconstruction or directed connectivity measures such as the Phase Slope Index or Granger causality would provide more precise neuroanatomical inference and allow assessment of information flow directionality within the networks identified here [60].

Despite these constraints, the convergence of behavioural, topographical, and connectivity evidence demonstrates that spatially congruent, embodied feedback fundamentally shapes both performance and neural dynamics in MTD-BCIs. The finding that users maintained functional control despite progressive assistance reduction, and that VR feedback sustained consistently higher DA than screen feedback across all training strategies and movement directions, establishes embodied spatial feedback as a key design principle for next-generation continuous BCIs targeting intuitive motor control and neurorehabilitation. Realising the full potential of this approach will require addressing the cross-session stability challenges identified here through co-adaptive frameworks that can accommodate evolving neural strategies alongside the richer feedback environments afforded by VR.

**Ethics Statement**

This study received ethical approval from the Ulster University Research Ethics Filter Committee, Faculty of Computing, Engineering and the Built Environment (reference: CEBE_RE-21-011-A). Ethical approval covered the full duration of the longitudinal data collection period (2021–2024). All experimental procedures involving human participants were conducted in accordance with the Declaration of Helsinki and institutional guidelines for research involving human subjects. Participants provided written informed consent prior to participation and were informed of the study procedures, data recording methods (including EEG, kinematic and eye-tracking data), their right to withdraw at any time without penalty, and the intended research use of anonymised data. All data were anonymised prior to analysis and dissemination. No personally identifiable information is included in this publication.

**Acknowledgements**

This work was supported by internal funds from the University of Bath and Ulster University; by access to the Tier 2 High Performance Computing resources provided by the Northern Ireland High Performance Computing (NI-HPC) facility, funded by the UK Engineering and Physical Sciences Research Council (EPSRC) under Grant number EP/T022175; the Spatial Computing and Neurotechnology Innovation Hub funded by the Department for Economy Northern Ireland and the UK Research and Innovation (UKRI) Turing AI Fellowship 2021-2026 funded by the EPSRC under Grant number EP/V025724/1.

**Competing Interests**

Prof Damien Coyle, Chief Investigator on this project, is the founder, Chief Executive Officer, and a shareholder of NeuroCONCISE Ltd., a company involved in the development of neurotechnology and wearable electroencephalography (EEG) systems.

**Data and Code Availability**

The raw multimodal dataset generated during this study (EEG, kinematics, eye-tracking, and psychological measures) will be made openly available via Zenodo (DOI: 10.5281/zenodo.16047021) upon publication. Analysis code and model implementations, including the ERSP-CNN-LSTM decoder, Ray Tune hyperparameter optimisation pipeline, and functional connectivity analysis, will be made available via linked GitHub repositories upon publication. A companion data descriptor paper describing the full dataset structure and acquisition protocol is currently in preparation. The ERSP-CNN-LSTM decoder and analysis pipelines are available from the corresponding author on reasonable request in the interim.